# Correlation between superconductivity and antiferromagnetism in Rb$_{0.8}$Fe$_2$Se$_{2-x}$Te$_x$ single crystals

Dachun Gu, Liling Sun*, Qi Wu, Chao Zhang, Jing Guo, Peiwen Gao, Yue Wu, Xiaoli Dong, Xi Dai and Zhongxian Zhao*

Institute of Physics and Beijing National Laboratory for Condensed Matter Physics, Chinese Academy of Sciences, Beijing 100190, China

We report the first experimental evidence for the intimate connection between superconductivity and antiferromagnetism in Rb$_{0.8}$Fe$_2$Se$_{2-x}$Te$_x$ single crystal under negative chemical pressure by substituting Se with isovalent Te atoms. Electrical resistance measurements in the temperature range from 4 K to 550 K demonstrate that both superconducting transition temperature (Tc) and Neel temperature ($T_N$) were suppressed continuously with the lattice expansion. When the Te concentration $x$ in Rb$_{0.8}$Fe$_2$Se$_{2-x}$Te$_x$ approaches 0.3, the superconducting transition temperature Tc is completely suppressed and the sample behaves like a semiconductor, meanwhile the characteristic peak of antiferromagnetic transition on resistance curve disappears. Our observation suggests that the pressure-induced lattice expansion can be used to tune the correlativity of superconductivity and antiferromagnetism.

*Corresponding author: llsun@aphy.iphy.ac.cn and zhxzhao@aphy.iphy.ac.cn

The discovery of superconductivity in LaFeAsO$_{1-x}$F$_x$ with Tc as high as 26 K [1] has attracted considerable attention in the world. During the past three years, several other types of iron-based superconductors with 122 structure (MFe$_2$As$_2$, M=Ca, Sr, Ba and Eu) [2-5], 111 structure (AFeAs, A=Li and Na) [6-7], 11 structure (FeSe) [8] and 42622 structure (Sr$_4$V$_2$O$_6$)Fe$_2$As$_2$) [9] were found. Among them, the highest Tc has reached 55K in SmFeAsO$_{1-x}$F$_x$ [10]. Recently, another new family of superconductors M$_x$Fe$_{2-y}$Se$_2$(M=K, Rb, Cs, or Tl substituted K, Rb) with Tc above 30 K and other unusual features were discovered [11-15], which simulated great interest in the community of condensed matter physics and material science. M$_x$Fe$_{2-y}$Se$_2$ superconductors show a number of peculiar features including electron-dominated carriers in the Fermi surface [16-19], high transition temperature of antiferromagnetic phase, a superstructure transition of Fe vacancies, and large ordering magnetic moments [20-21]. Application of internal or external positive pressure on these superconductors showed that the superconducting transition temperature was suppressed with increasing pressure [22-24]. In this work, we investigate the effect of negative pressure by substituting selenium (Se) with isovalent tellurium (Te) in Rb$_{0.8}$Fe$_2$Se$_2$. As in the case of positive pressure, we found that the superconducting critical temperature is suppressed with an increment of Te concentration. To our surprise, we find that both the superconducting and antiferromagnetic orders vanish at the same time when the doping concentration of Te approaches to 0.3, indicating the inherent relationship between superconductivity and antiferromagnetism in this material. By comparison with positive pressure effect, we propose that pressure-free sample (neither positive nor negative pressure) possesses an optimal lattice configuration in favor of the highest transition temperature of superconductivity.

Single crystals of $Rb_{0.8}Fe_2Se_{2-x}Te_x$ were grown by a self-flux method with several steps. First, precursors of FeSe and FeTe were synthesized by solid reaction method. High purity Fe, Se and Te powder were mixed together with nominal composition of Fe:Se and Fe:Te in a mortar. The mixture was put in a furnace and heated to 700 ℃ with a rate of 100℃/hour, kept at this temperature for 24 hours and then cooled down to room temperature naturally. Second, the precursors were mingled with Rb in a glovebox and loaded into an alumina crucible, then sealed in an evacuated silica ampoule. Third, the sealed silica tube was placed in a furnace, slowly heated to 1000℃ and kept at this temperature for 5 hours, afterward heated up to 1100℃ and kept for another 5 hours. Finally, the samples were cooled down to 800℃ with a rate of 4℃/hour, followed by shutting off the power of the furnace.

The resulting samples were characterized by x-ray diffractometer with Cu K$\alpha$ radiation ($\lambda$=1.5418Å) in the 2$\theta$ scan mode, as shown in Fig 1a. Sharp (00l) peaks reflect that the sample orientates well in the *c* direction. Electrical resistance and magnetic susceptibility measurements below 300 K were performed with physical property measurement system (PPMS-9) and superconducting quantum interference device (SQUID-XL1). As shown in Fig.1b and c, a clear superconducting transition at 31.4K determined by the resistance measurement and at 30K detected by magnetic susceptibility measurement was observed, indicating a bulk nature of superconductivity in $Rb_{0.8}Fe_{1.8}Se_2$.

To investigate how the tellurium (Te) substitution influences the lattice distortion, we first performed x-ray diffraction measurements at room temperature for $Rb_{0.8}Fe_2Se_{2-x}Te_x$. As shown in Fig.2a, all samples can be well indexed to tetragonal structure, demonstrating that no phase transition occurs in the substitution range investigated. However, the lattice expansion was clearly observed with increasing Te concentration (Fig.2b) due to the substitution of tellurium whose ionic radius of Te is larger than that of Se. The expansion in the $c$ direction (1.3%) is slightly larger than that in the $a$ direction (0.7%). Fig.2c shows how the volume of a unit cell grows with increasing Te concentration. It confirms that Te substitution induces a negative pressure, as expected.

Figure 3 shows the temperature dependence of electrical resistance for four samples with different Te concentrations. We found that with increasing negative pressure, superconducting transition temperature (Tc) of the sample with $x$ smaller than 0.25 was suppressed (Fig.3a). For the sample with higher Te concentration ($x \geq 0.3$), the superconductivity is fully destroyed and its resistance exhibits semiconducting behavior (Fig.3b). The response of superconductivity to negative pressure in Te-doped $Rb_{0.8}Fe_2Se_{2-x}Te_x$ is similar to that to positive pressure in S-doped $K_xFe_{2-y}Se_2$ or compressed $K_{0.8}Fe_{1.7}Se_2$ [22-24]. Figure 4 compares the pressure dependences of Tc and the sample volume in the positive and negative pressure cases.

Previous studies on FeAs-based superconductors showed that the superconductivity is very sensitive to lattice distortion [25]. This distortion consists of the change in As-Fe-As angle [26] and the anion height [27] which can be produced either by chemical or physical pressure [28]. To inspect the effect of the lattice distortion on observed suppression of superconductivity in $Rb_{0.8}Fe_2Se_{2-x}Te_x$, we computed the Se-Fe-Se angle and the anion height, based on XRD data, together with Tc as a function of Te concentration (negative pressure). As shown in Fig.5, a negative pressure leads to an enhancement of anion height and a reduction of the Se-Fe-Se angle. As the anion height increases or the bond angle decreases, the superconducting temperature goes down. At x=0.3, where the anion height and the bond angle change their trend, the superconducting transition of $Rb_{0.8}Fe_2Se_{2-x}Te_x$ is fully suppressed. Our results indicated that the superconductivity of $Rb_{0.8}Fe_2Se_{2-x}Te_x$ is governed by the level of their lattice distortion.

To identify the effect of negative pressure on the long-range order of antiferromagnetism and the formation of Fe-vacancies in high temperature regime, which are related to superconductivity of the iron chalcogenides, we measured electrical resistance from 300 to 550 K for $Rb_{0.8}Fe_2Se_{2-x}Te_x$ single crystals. The resistance and its derivate as a function of temperature were plotted in Fig.6. There are two dips on the dR/dT curve. One is associated with the formation of Fe-vacancy order (whose transition temperature is defined as $T_S$), the other is related to the antiferromagnetic phase transition (whose transition temperature is defined as $T_N$) [29]. For Te-free sample, we found that its $T_S$ and $T_N$ are about 536 K and 493K respectively, which are slightly lower than the corresponding values reported in Ref. [29]. This may be due to the subtle difference of samples. Both $T_N$ and $T_S$ decrease with increasing Te concentration. This indicates that lattice expansion favors the formation of Fe vacancies as well as the transition of paramagnetic-to-antiferromagnetic phase. Surprisingly, we found that $T_N$ vanishes at $x \geq 0.3$ where the sample lost its superconductivity. Figure 7 shows the phase diagrams of Tc (P) and $T_N$ (P). The simultaneous disappearance of Tc and $T_N$ in $Rb_{0.8}Fe_2Se_{2-x}Te_x$ suggests that the superconductivity is correlated with antiferromagnetism. Recently, μsR, Mossbauer, neutron scattering and Raman experiments also showed that the superconducting long range order in $MFe_2Se_2$ coexists with the antiferromagnetic long range order [20, 29-32]. The superconducting pairing mechanism in this intercalated $MFe_2Se_2$ superconductor remains an open question.

In summary, we show that the partial substitution of selenium by tellurium atoms can create negative pressure which leads to the lattice expansion. Transport measurements suggest that superconductivity and antiferromagnetism of $Rb_{0.8}Fe_2Se_{2-x}T_x$ single crystals are suppressed by the Te substitution. At x= 0.3, both the superconducting and antiferromagnetic orders are completely destroyed and the sample becomes semiconducting. This indicates that the superconductivity in $Rb_{0.8}Fe_2Se_{2-x}Te_x$ is correlated to its antiferromagnetism. Furthermore, our results indicate that the superconductivity in $Rb_{0.8}Fe_2Se_{2-x}Te_x$ is intimately related to the level of their lattice distortion. Comparing with the effect of positive pressure, application of negative pressure is also unfavorable to achieve the optimal value for superconductivity of $Rb_{0.8}Fe_2Se_{2-x}Te_x$ (x<0.3).

Acknowledgement

We would like to thank P. C. Dai, T. Xiang，X. J. Chen and Z. Fang for useful discussions. This work was supported by NSCF (10874230 and 11074294), 973 projects (2010CB923000 and 2011CBA00109), and Chinese Academy of Sciences.

Figure captions:

Fig.1 (a) (Color online) X-ray diffraction (XRD) pattern of $Rb_{0.8}Fe_2Se_2$ single crystal, displaying only (00l) peaks. (b) Temperature dependence of electrical resistance. (c) Magnetic susceptibility as a function of temperature for zero-field-cooling and field-cooling processes.

Fig.2 (a) (Color online) X-ray diffraction patterns of powdered $Rb_{0.8}Fe_2Se_{2-x}Te_x$ which is ground from the single crystals. (b) Lattice parameters of $Rb_{0.8}Fe_2Se_{2-x}Te_x$ as a function of Te concentration. (c) Te substitution dependence of volume, showing that Te substitution expands the lattice.

Fig.3 (a) (Color online) Temperature dependence of resistance for $Rb_{0.8}Fe_2Se_{2-x}Te_x$ (x=0~0.25) at zero magnetic field. The inset displays the decrease of Tc with Te substitution. (b) Resistance as a function of temperature for $Rb_{0.8}Fe_2Se_{2-x}Te_x$ (x ≥0.3), showing typical semiconducting behavior.

Fig.4 (Color online) (a) and (c) Tc and volume as a function of S concentration in $K_xFe_{2-y}Se_{2-z}S_z$, data in (a) and (c) are from Ref. 24. (b) and (d) as a function of Te concentration in $Rb_{0.8}Fe_2Se_{2-x}Te_x$.

Fig.5 (Color online) (a) Se-Fe-Se angle dependence of Te concentration in $Rb_{0.8}Fe_2Se_{2-x}Te_x$. (b) Anion height from Fe layer as a function of Te concentration.

Fig. 6 (Color online) (a) High temperature resistance for $Rb_{0.8}Fe_2Se_{2-x}Te_x$ (x=0~0.4) from 250K to 550K. (b) The first derivation of the temperature resistance, exhibiting two important transitions.

Fig.7 (Color online) Antiferromagnetic and superconducting phase diagram of $Rb_{0.8}Fe_2Se_{2-x}Te_x$ single crystals.

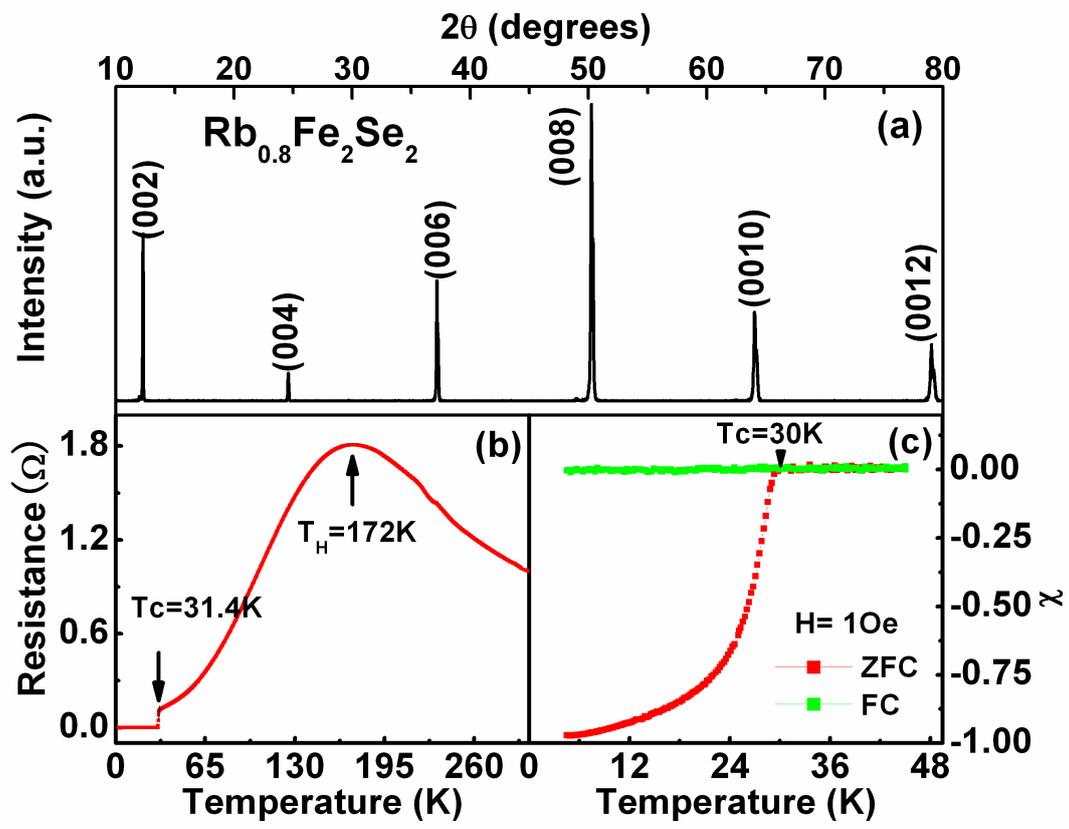

Fig.1 Gu et al

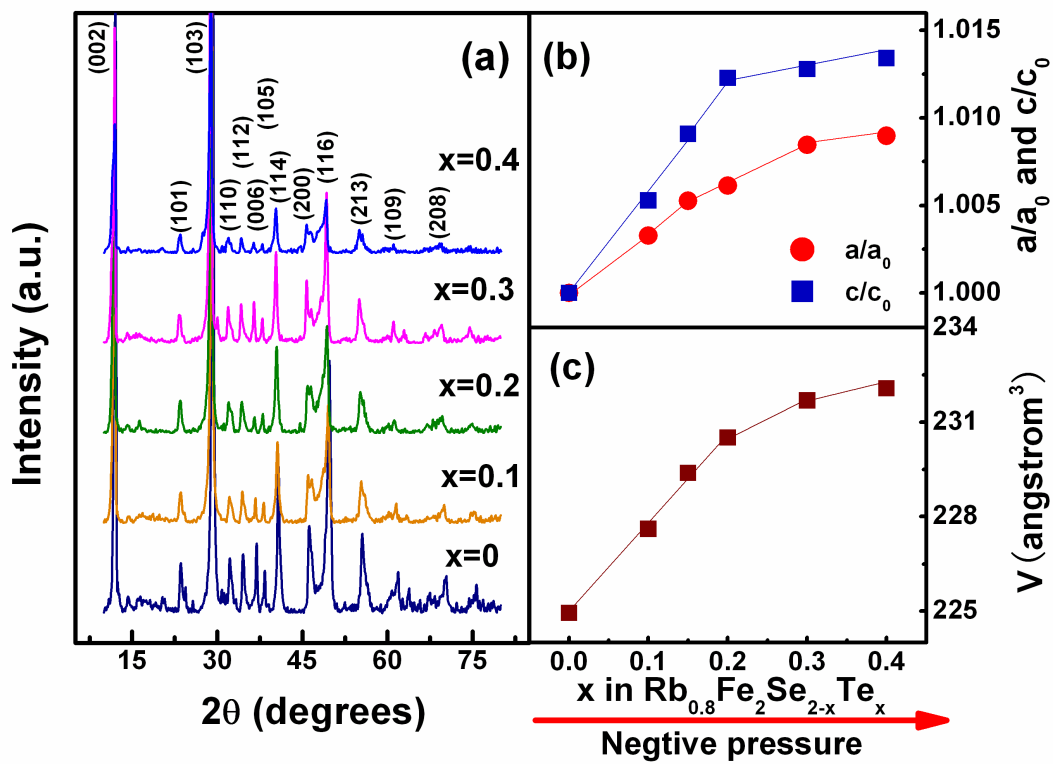

Fig.2 Gu et al

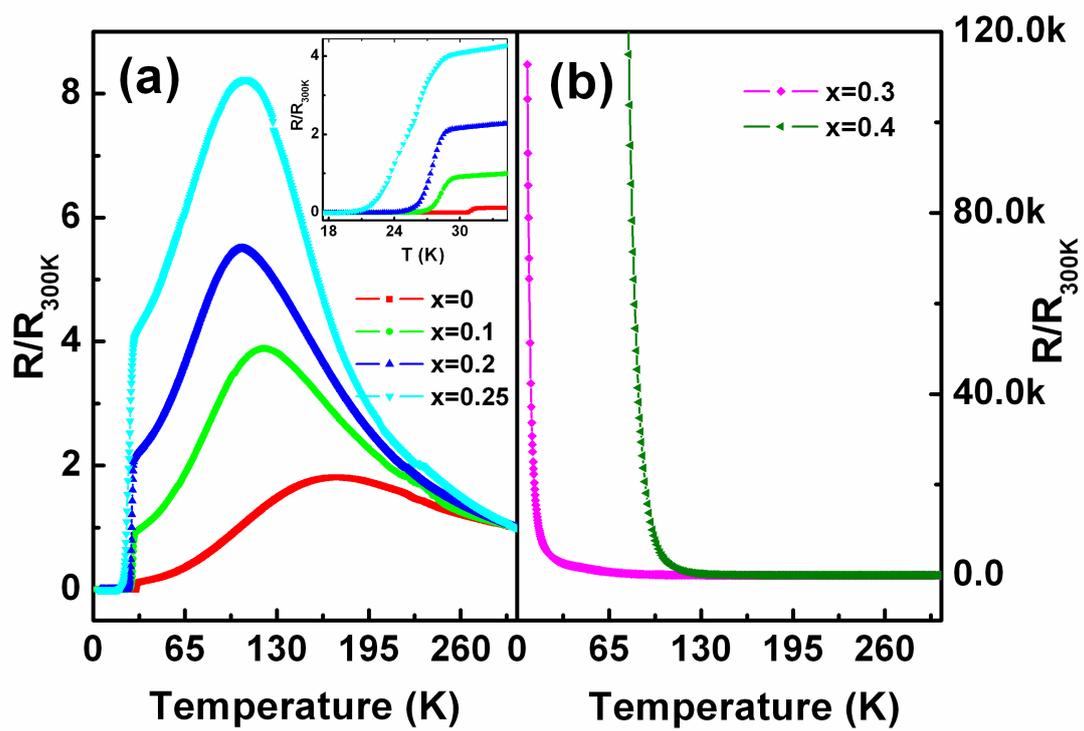

Fig.3 Gu et al

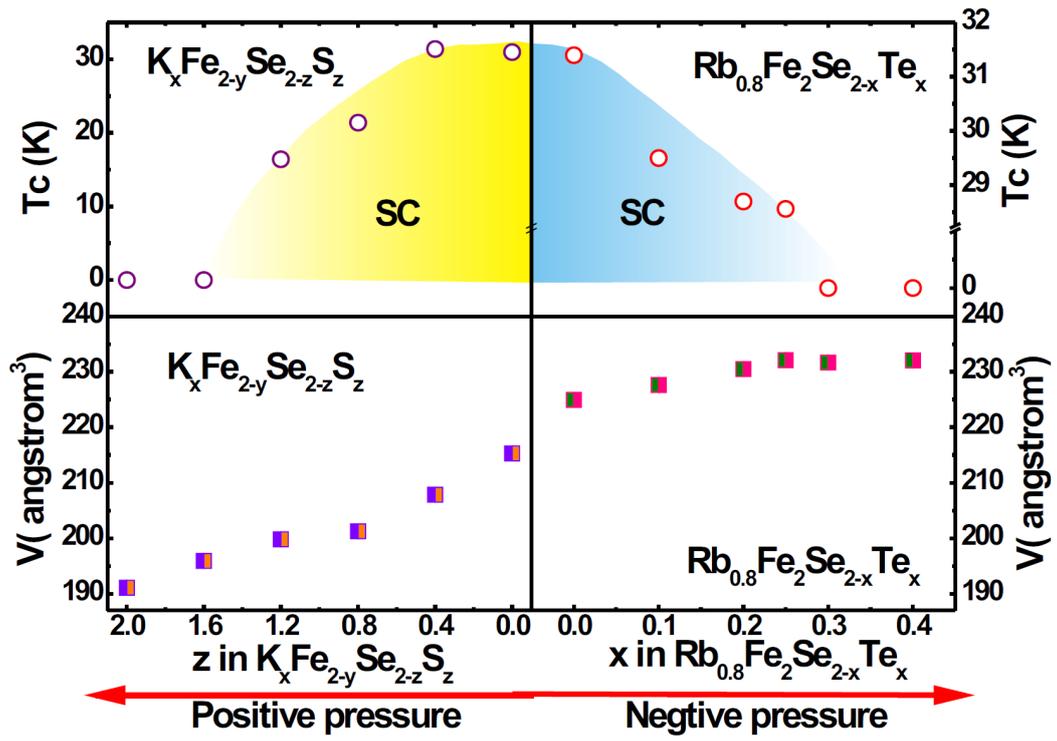

Fig.4 Gu et al

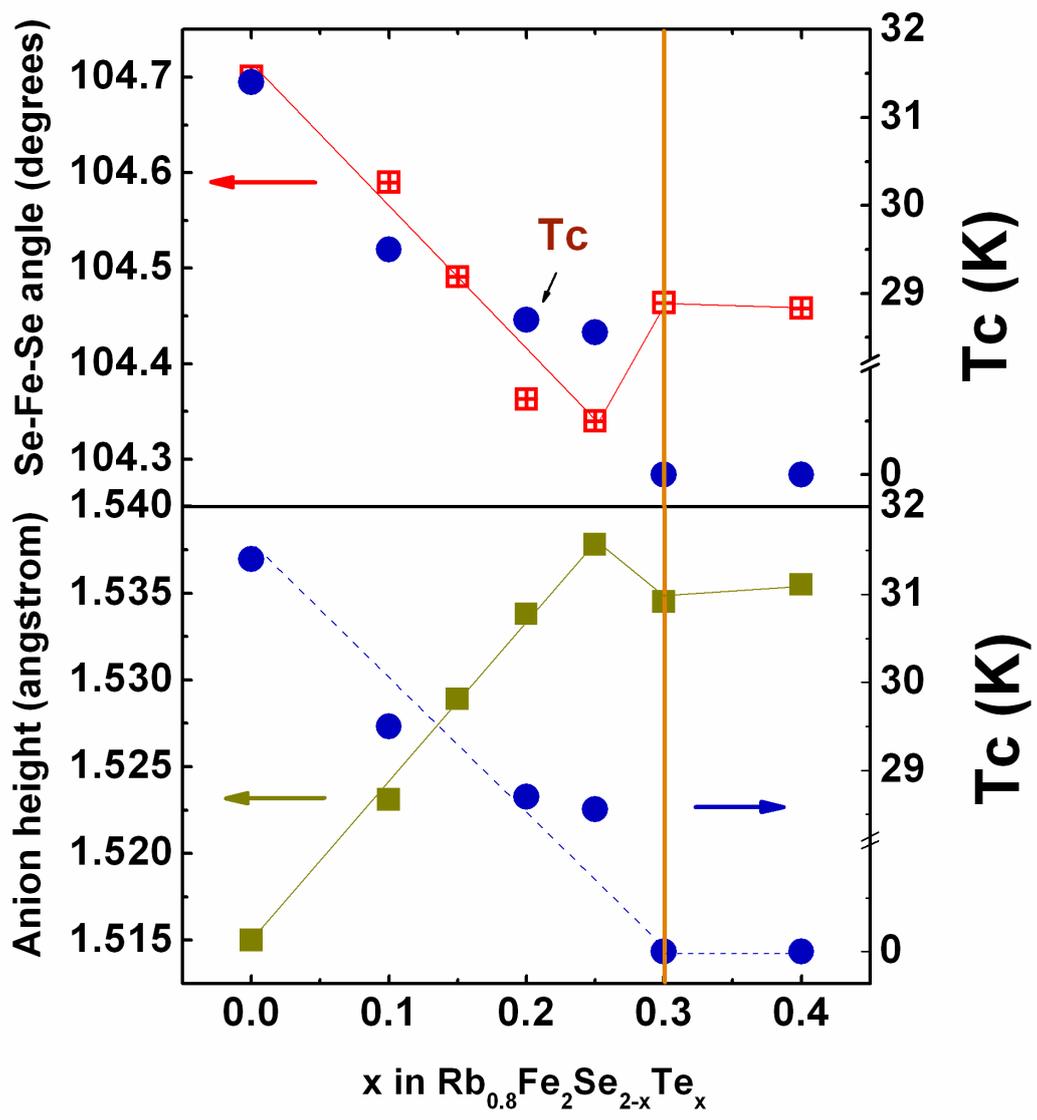

Fig.5 Gu et al

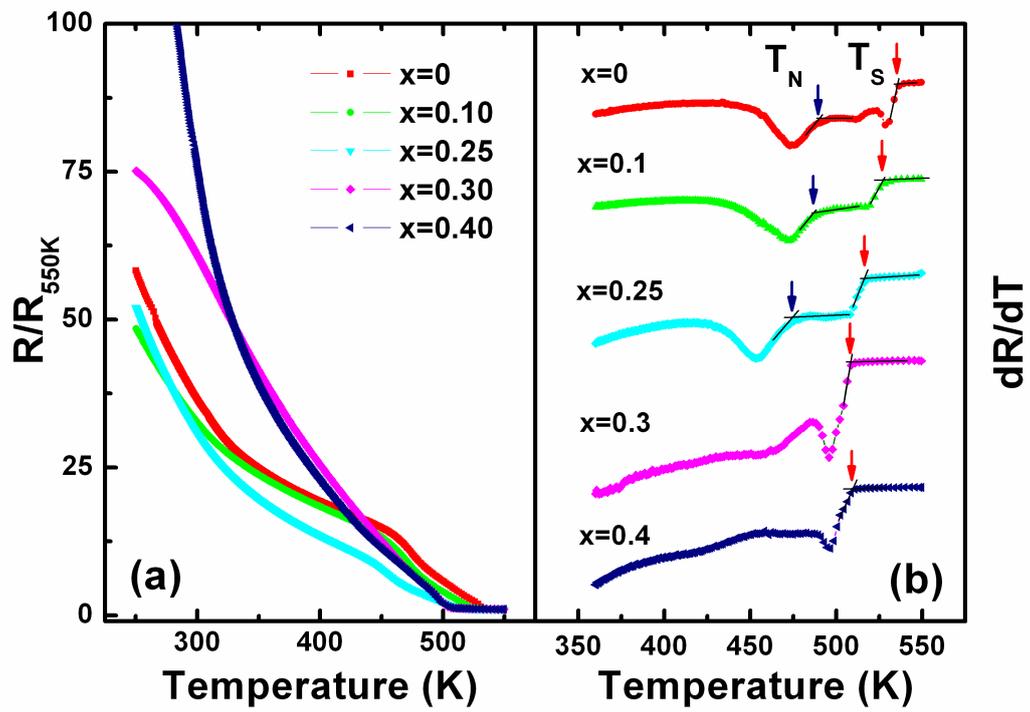

Fig.6 Gu et al

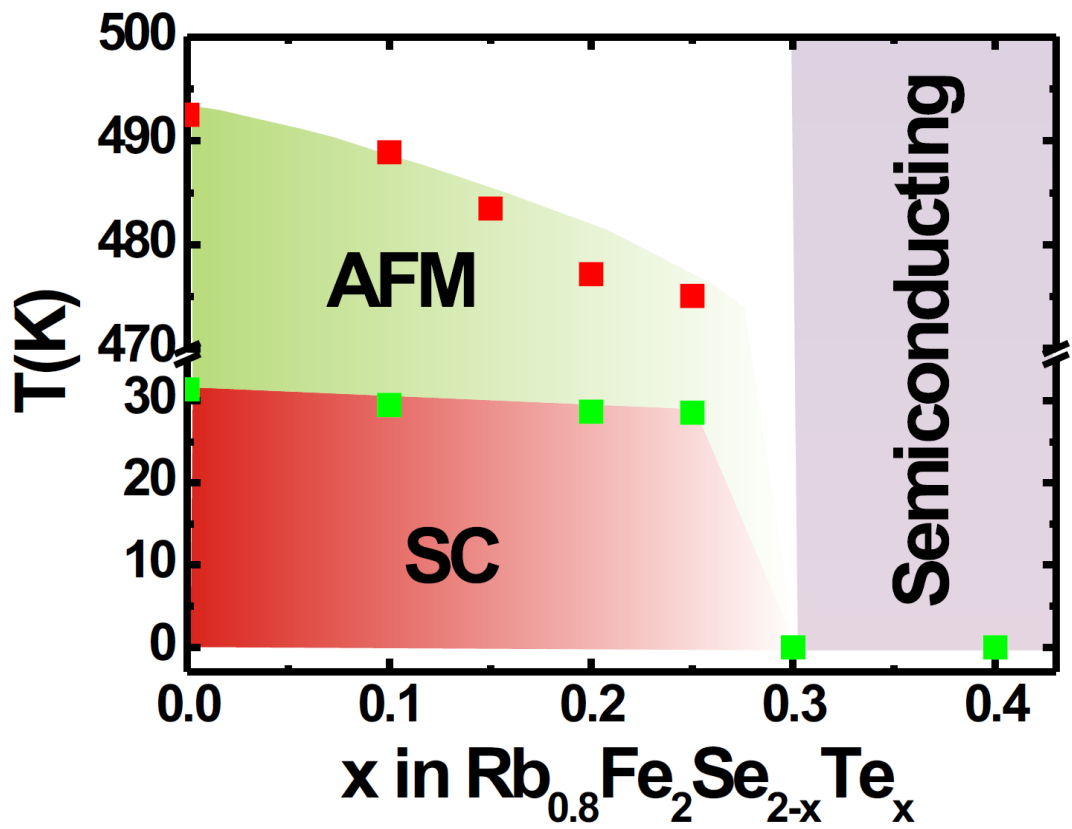

Fig.7 Gu et al